# Level of Scientific Readiness with Ternary Data Types

Eldar Knar

Fellow of the Royal Asiatic Society of Great Britain and Ireland

[https://orcid.org/0000-0002-7490-8375](https://orcid.org/0000-0002-7490-8375)

eldarknar@gmail.com

**Abstract**

In addition to the technology readiness level (TRL the scientific readiness level (SRL) has been introduced as a more authentic and adequate tool for determining the *status quo* of scientific and scientific-technical projects of fundamental or applied nature.

The SRL includes 10 levels of scientific readiness, namely, the FRL (Fundamental Readiness Level), the ARL (Applied Readiness Level), and the IRL (Innovation Readiness Level).

For quantitative and visual assessment of the level of scientific readiness, a system of positional ternary (three-valued, trinary) codes with integer trits (ternary digit) is introduced. The ternary code interprets the degree of project elaboration according to the fundamental, applied, and innovation vectors/trits FRL/ARL/IRL. Examples of scientific project assessments are provided.

The main characteristics of the new scale and assessment of the level of readiness and the application areas of the scientific readiness level are noted. Tasks for further development of the system of scientific readiness levels are formulated.

**Acknowledgments and Declaration:** This article was prepared within the programme Trust Fund Study OR 11465474 "Scientific foundations of modernization of the education system and science" (2021-2023, Altynsarin National Academy of Education, Astana, Republic of Kazakhstan).

No AI platforms (LLM, such as ChatGPT) were used in writing the manuscript. This is the principled position of the author.

# 1. Introduction

The technology readiness (TRL) level, including the MRLs and CRLs, according to NASA (the National Aeronautics Space Administration (NASA), the US Department of Defense) {NASA, 2019, nasa.gov] and the European Union (DLR, Germany, CNES, France, ESA - European Space Agency) [EU Commission Decision, 2014], is widely used for assessing technologies and technological solutions and as one of the criteria for project selection or evaluation.

In NASA notation, the TRL is interpreted through nine readiness levels depending on the parameters of space technologies:

**Table 1.** Full version of the TRL in the original NASA notation

| TRL | DEFINITION | EQUIPMENT DESCRIPTION | SOFTWARE DESCRIPTION | EXIT CRITERIA |
|---|---|---|---|---|
| 1 | Observe and report on basic principles | Scientific knowledge generated from hardware technology concepts/applications | Scientific knowledge generated based on basic software architecture properties and mathematical formulation | Peer-reviewed publication of research underlying based on the proposed concept/application |
| 2 | The concept and/or application of the technology has been formulated | An invention begins, a practical application is identified but is speculative, there is no experimental confirmation or detailed analysis to support the conjecture. | Practical applications have been identified, but are speculative and there is no experimental confirmation or detailed analysis to support the hypothesis. The main properties of algorithms, representations and concepts are defined. Basic principles are encoded. Experiments were carried out with synthetic data | Documented description of the application/concept addressing feasibility and benefit issues |
| 3 | Analytical and experimental proof of critical functions and/or characteristics of the concept | Analytical studies place the technology in context, and laboratory demonstrations, modeling and simulation validate analytical predictions | Develop limited functionality to validate critical properties and predictions using nonintegrated software components | Documented analytical/experimental results supporting key parameter predictions |
| 4 | Validate a component and/or layout in a lab environment | Created and operated a low fidelity system/component mockup to demonstrate core functionality and critical test environments, and defined associated performance environments relative to the final production environment. | Core functionally important software components are integrated and functionally validated to establish compatibility and begin architecture development. Relevant environments are identified and performance in that environment is predicted | Test performance is documented to demonstrate consistency with analytical predictions. Documented definition of the appropriate environment |
| 5 | Validate the component and/or layout in the appropriate environment | Created and operates a medium fidelity system/component mockup to demonstrate overall performance in a simulated operating environment with realistic support elements demonstrating overall performance in critical areas. Performance predictions are made for subsequent stages of development | Implemented and interfaced software elements end-to-end with existing systems/simulations to suit the target environment. A prototype system tested in an appropriate environment consistent with predicted performance. Predicting performance in the operating environment. Prototype | Documented test performance demonstrating consistency with analytical predictions. Documented definition of scaling requirements |
| | | | implementations developed | |
| 6 | Demonstration of a model or prototype of a system/subsystem in an operating environment | Built and operates a high fidelity system/component prototype that adequately addresses all critical scaling issues and operates in an appropriate environment to demonstrate operations under critical environmental conditions | Prototype software implementations are demonstrated on full-scale, realistic problems. Partial integration with existing hardware and software systems. Limited documentation available. Engineering feasibility fully demonstrated | Documented test performance demonstrating consistency with analytical predictions |
| 7 | Demonstration of a prototype system in an operating environment | Created and operated a high-fidelity engineering image that adequately addresses all critical scaling issues and runs in the appropriate environment to demonstrate performance in a real-world operating environment and on the appropriate platform (ground, air, or space) | Prototype software exists containing all key features available for demonstration and testing. Well integrated with operational hardware and software systems, demonstrating operational feasibility. Most software bugs have been fixed. Limited documentation available | Documented test performance demonstrating consistency with analytical predictions |
| 8 | The system is actually complete and "flight tested" through testing and demonstration. | The final product in its final configuration is successfully demonstrated through testing and analysis for the intended operating environment and platform (ground, air or space). | All software is fully debugged and fully integrated with all operating hardware and software systems. All user documents, training documents and service documentation are complete. All functionality is | Test performance documented to support analytical predictions |

| | | | successfully demonstrated in simulated operating scenarios. Verification and Validation (V&V) completed | |
|---|---|---|---|---|
| 9 | Current flight of the system, confirmed by successful mission operations. | The final product is successfully operated in a real mission. | All software is fully debugged and fully integrated with all hardware and software systems. All documents have been completed. Sustainable engineering support for the software is provided. The system is successfully operated in the operating environment | Recorded mission operational results |

In the Kazakhstani scientific system, the practice of TRL for evaluating and expertizing the level of readiness of scientific and research-intensive projects is also introduced at the official level. This proposal was actualized by the Minister of Science and Higher Education, Sayasat Nurbeck: "1. Implement the practice of determining the level of technological readiness (TRL) of ideas and developments of universities, scientific organizations and startup companies" [https://primeminister.kz, 14.02.2023].

In this sense, with the development of the new Law "On Science and Technology Policy" of the Republic of Kazakhstan, issues of scientific expertise are being actualized, especially in the field of pure and related fundamental and applied research

In particular, the official "Methodology for determining the level of technological readiness (TRL) of scientific organizations and research universities, and their developments" has been adopted (Appendix to the order of the Chairperson of the Science Committee of the Ministry of Science and Higher Education of the Republic of Kazakhstan of "July 18, 2023 No. 112-nzh).

According to the Methodology (quote): "types of scientific and scientific-technical activities mean scientific research and development classified as fundamental and applied scientific research, and experimental developments. The type of production activity refers to the stage of production development, including pilot production and certification, as well as serial production" (end of quote).

The hierarchy of technology readiness levels (TRLs) in the methodology is as follows (according to "Description of technology readiness levels and their comparison with types of activities. Appendix to the Methodology for Determining the Level of Technological Readiness (TRL) of Scientific Organizations and Research University and Their Developments":

Level 0 (?!) - Fundamental research
Level 1 - Fundamental research
Level 2 - Applied research
Level 3 - Applied research
Level 4 - Experimental developments
Level 5 - Experimental developments
Level 6 - Experimental developments
Level 7 - Experimental developments
Level 8 - Pilot production and certification
Level 9 - Production

It is clear that this methodology does not reflect the uniqueness, peculiarity, or specificity of purely scientific research. In particular, science is not limited to "pilot production and certification" or "production".

Let me give you a simple example—fundamental research. According to this hierarchy, the overwhelming majority of fundamental research will have a level of technological readiness equal to 0 or 1.

In other words, a TRL of zero or one is the ceiling for half of scientific research. This circumstance alone does not inspire optimism. In other words, you will remain zero no matter how far you advance in your purely fundamental and exploratory research.

In general, if you are engaged in fundamental research, boldly write TRL 0 or 1. You will not go wrong.

Of course, this scale of technology readiness levels applies to scientific projects that can be commercialized. However, even commercializable scientific developments may not be limited to "pilot production" or "production" in the literal sense. However, they may still have commercial success or market value.

Therefore, this is not only a fundamental question but also a terminological and methodological one. Those who apply or will apply this methodology may understand it too literally. In other words, if the Methodology says "fundamental and applied", it will be applied to all scientific research without exception.

For example, in the new "Competition Documentation for Grant Financing of the Most Promising Projects for Commercializing the Results of Scientific and (or) Scientific-Technical Activities for 2024-2026" (approved by the Decision of the Board of JSC "Science Foundation" of December 7, 2023 No. 48, Astana, 2023), the level of technological readiness is already introduced as a requirement (quote): "the results of scientific and (or) scientific-technical activities at the time of application must correspond to the level of technological readiness (TRL) 6 (six) or higher" (end of quote).

If this requirement is introduced to all other grant programs (without commercialization) in a literal "fundamental or applied" understanding, then how it will look in practice is not entirely clear.

Therefore, at the level of readiness of scientific research, the form (direction and nature of science) should not be important, but the content—the staginess within the research itself. It does not matter whether the research is applied or fundamental. In other words, to improve the quality and effectiveness of scientific research, it is necessary to transition the system of expert assessments from the paradigm of "form and content" to the paradigm of "content".

Therefore, for purely scientific projects, the level of technology readiness (technology readiness level), including the MRLs (manufacturing readiness levels) and CRLs (commercialization readiness levels) based on the TRA-technology readiness assessment paradigm, is not applicable in its pure form or has significant limitations in applicability.

Accordingly, it is necessary to create a universal and standardized methodology for assessing the scientific readiness of fundamental and applied R&D and applied R&D for practical application in the scientific and grant system of the Republic of Kazakhstan. In particular, management decisions in the national scientific system should be made.

## 2. Scientific Readiness Level (SRL)

A new model of transformation is needed through the creation of a system for assessing the readiness level of technologies authentic to purely scientific fundamental and applied projects with universalization for all branches and directions of scientific research. Therefore, it is advisable to replace the readiness level scale in the notation of NASA and the European Union with a new parameter,

SRL—the Scientific Readiness Level—which reflects the specificity of R&D and applied scientific projects rather than purely technological products. The development, justification, and elaboration of this standardized new system of scale and assessment of the readiness level of scientific projects and research are necessary for practical application in the scientific and grant system of the Republic of Kazakhstan.

Previously, various modifications of the technology readiness level (TRL) were proposed. However, they did not go beyond the scope of the technological readiness level itself. In our case, we are not talking about modifications to the TRL but about a new scale and assessment specifically for purely scientific fundamental and applied projects and research.

Previously, the scientific readiness level was interpreted through the definition of SciRL with three readiness levels: SciRL1, SciRL2, and SciRL3. The methodology of scientific readiness, interpreted as SRL (Scientific Readiness Levels), has been developed and specified through 10 levels of SRL with structural components FRL, ARL, and IRL.

The SRL readiness level includes three components based on the nature of the basic or grant project proposal, namely, FRL/ARL/IRL:

*FRL - Fundamental Readiness Level*
*ARL - Applied Readiness Level*
*IRL – Innovation Readiness Level*

The conceptual initial definitions of the scientific readiness level are reflected in the Table:

**Table 2**. Structure of Scientific Readiness Levels

| | SRL LEVELS for *R&D* | | |
|---|---|---|---|
| **LEVEL** | **FRL**<br>**FUNDAMENTAL**<br>**READINESS** | **A.R.L.**<br>**APPLIED READINESS** | **IRL**<br>**INNOVATION**<br>**READINESS** |
| 10 | Further development of the scientific project, scaling an idea, creating a scientific school, followers, recognition, insignia | Further development of a scientific, scientific-technical or innovative project, recognition, insignia, scaling, mass | agreement, contract, investment, acquisition of license, Scaling, entering markets, further modernization and |

| | | access to consumers or customers | rationalization, mass market, mass demand |
|---|---|---|---|
| 9 | Research results are recognized and cited (scientometrics), insignia (awards) | Research results are recognized and cited (scientometrics, memorandums, agreements of intent) | Patents, copyrights or licenses acquired by third parties or organizations, partnership agreement, (co)investment agreement |
| 8 | The content is published (including electronic versions of books in the registry) and indexed in the SCOPUS, WoS database or published in the KKSON list | The content is published (including electronic versions of books in the registry) and indexed in the SCOPUS, WoS database or published in the list of KKSON (Committee for Quality Assurance in Education of the Ministry of Education of the Republic of Kazakhstan), a ready-made technology or product. Patent or legal documents | Finished product, license, product, patent, copyright, access to an individual consumer, agreement, memorandum |
| 7 | Content (article, book, abstract, plenary report, poster presentation) sent to the relevant editorial office printing house or organizing committee (active mode) | Content (article, book, theses, plenary report, poster presentation) has been sent to the relevant editorial office, printing house or organizing committee | A patent, copyright, or other law enforcement document has been applied for |
| 6 | Ready article for publication (passive mode) | Ready material for publication | Industrial prototype, working model, |
| 5 | Research in process (proactive mode without funding) | Research in progress (proactive mode without funding), sample, working model | Business plan, sample, working model (at the process stage), bench model |

| 4 | Availability or understanding of research methodology | Availability or understanding of research methodology or technology | Implementation technology, including logistics, suppliers, consumers, pricing policy |
|---|---|---|---|
| 3 | Approbation among colleagues, report at a seminar or council | Approbation at the academic council, among colleagues, report at a seminar or council, reference search | Patent search, databases |
| 2 | Bibliographic analysis (relevance, necessity, significance) | Bibliographic analysis, marketing analysis | Marketing analysis (comparison, options, application) |
| 1 | Hypothesis, idea | Idea, need | Idea, need, offer, trend, situation |
| 🟨 | *Results will be achieved upon implementation of the project* | | |
| 🟩 | *Parameter level reached* | | |

In essence, the scientific definitions of FRL/ARL/IRL are not entirely the same as the original definitions of TRL/MRL/CRL. However, they include technological, production, and market components, taking into account the specifics of the scientific sphere, methodology, and organization of scientific work.

This SRL scale can be adjusted, supplemented, edited, or compiled. However, in general, it reflects the readiness level of knowledge-intensive projects. The stages do not necessarily have to be sequential. Moreover, the value of readiness is determined by the maximum or average value among the implemented components.

If a project has a readiness level of 10 (meaning all previous stages 1-9 have been completed) according to SRL, then the project is a priority for grant funding without additional conditions (e.g., private coinvestment).

If a project has a readiness level of 9 (meaning all previous stages 1-8 have been completed) according to SRL, then the project is a priority for grant funding (if further development is required) with additional conditions (e.g., private coinvestment).

For the remaining stages, the priority of funding is identical and authentic. It all depends on the assessment of the idea itself, and authenticity eliminates obstacles to the development path and the emergence of new ideas, new directions, and project performers (new scientific workers, researchers, and innovators).

However, it should be noted that SRL (similar to TRL) reflects only the readiness level, not the level of scientific novelty, potential, or perspective of a scientific idea or project. In other words, SRL reflects the stages of the project but not its scientific content.

The passion of the scientific readiness level is interpreted as follows:

**Table 3:** Passport System and Methodology of Scientific Readiness Levels

| No. | CRITERIA | DEFINITIONS, OBJECTS, ELEMENTS |
|---|---|---|
| 1 | Application area | Scientific policy, management of the scientific system, reporting of R&D and RSTTD, grant programs, examination of scientific projects |
| 2 | Terms and Definitions | SRL - Scientific Readiness Levels, levels scientific readiness <br> FRL - Fundamental Readiness Level, levels fundamental readiness , <br> ARL - Apllied Readiness Level, levels applied readiness , <br> IRL – Innovation Readiness Level, levels of innovative readiness |
| 3 | Functions (purpose) | Authenticity and adequacy of determining the scientific readiness of projects and studies within the framework of R&D and RNNTD (instead of or in addition to TRL) |
| 4 | Target Operations | sequence of states or events authentic for performing (achieving) a specific level within the framework of scientific processing |
| 5 | Critical elements | Expertise <br> Grade <br> Publication <br> Conference <br> Seminar <br> Scientific report <br> Poster presentation <br> Plenary report <br> Abstracts <br> Book/monograph <br> Patent for invention <br> Certificate of authorship <br> Agreement |

| | | Grant Agreement<br>Business plan/master plan<br>Certificate |
|---|---|---|
| 6 | Process | The hierarchy of levels can be adapted to the specifics of individual R&D and RNSTD<br>The sequence of levels can be end-to-end<br>The beginning of the next level does not necessarily follow from the previous one<br>Starting the next level does not always mean finishing the previous one |
| 7 | Validation | Review by the Academic Council<br>Review by expert<br>Content for review<br>OIS (author's floating object) at registration<br>Sent to the editor<br>Sent to the organizing committee |
| 8 | Verification | Insignia<br>Academic degree/rank<br>Grant<br>Project-targeted financing<br>Collaboration<br>Positive decision of the Academic Council<br>Positive expert opinion<br>Positive review<br>High level of originality (anti-plagiarism, certificate)<br>SCOPUS/WoS data<br>Acceptance for publication<br>Publication by quartiles and percentiles<br>Book/monograph (ISBN , indexing) |

## 3. Assessment based on ternary encoding

As a form of assessment for SRL (scientific readiness level), a positional ternary code is recommended: N.N.N., where N represents integer trit. For example:

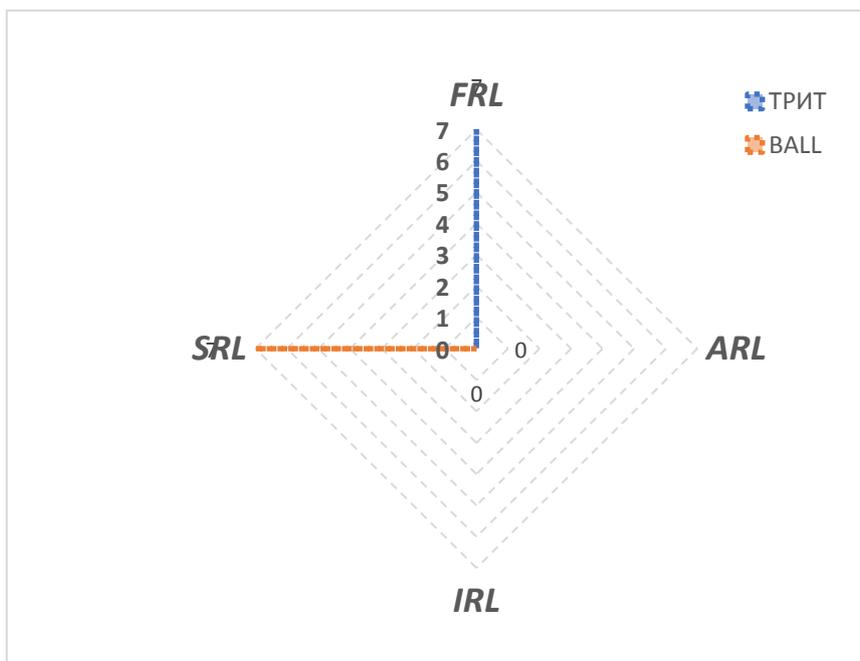

**Fig 1.** 7.0.0 (converted score 7) – The project is purely fundamental research. It is at the stage of developing scientific content in an active mode, possibly at the stage of submitting the article to the editor, undergoing peer review, or being accepted for publication.

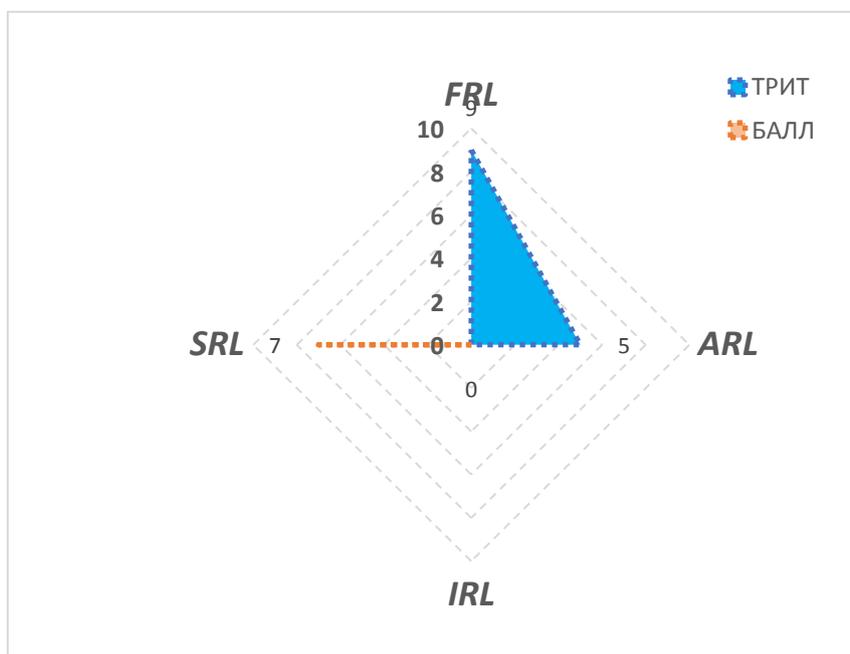

**Fig 2.** 9.5.0 (converted score 7) – This project combines fundamental research with applied value. The results are acknowledged and cited. The applied aspect is in the stage of research, model development, or sample creation.

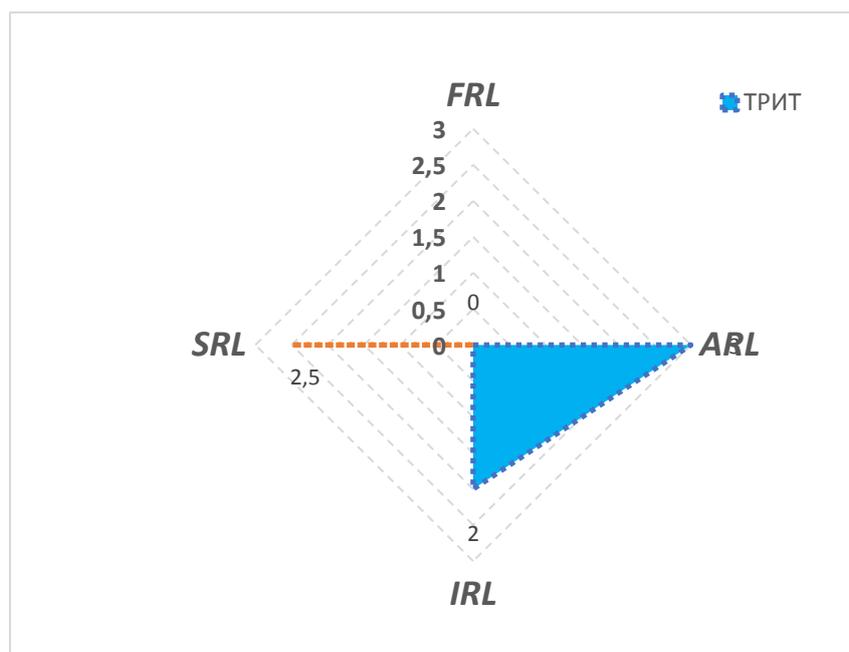

**Fig 3.** 0.3.2 (converted score 2.5) – The project has market potential. It has been tested in a relevant scientific environment and approved. It is currently undergoing a patent search for market entry (commercialization).

The last parameter in the ternary code, with a nonzero argument/trit, determines the specific sphere and direction of the scientific project. The ternary code reflects the degree of elaboration of the project along the fundamental, applied, and innovative vectors/trits FRL/ARL/IRL. The average score reflects the overall level of elaboration of the scientifically ready project.

The fundamental difference between SRL and TRL lies in the fact that SRL takes into account the specifics, nuances, and intricacies of scientific processing inherent in purely fundamental and applied scientific research. In this sense, SRL more adequately and authentically assesses the readiness level of a scientific project than does TRL. The application of the TRL helps avoid errors, inaccuracies, and discrepancies in assessing the readiness level, which may arise when the TRL is directly used.

It should be noted that the correct and appropriate choice of assessment methodology and methodology for evaluating the condition, quality, and effectiveness of an object is the most important factor in the expert evaluation of scientific research. Furthermore, the use of SRL as a standard for assessing the readiness level of scientific research does not preclude the corresponding use of TRL. They can complement each other in various combinations or proportions in the managerial or expert assessment of scientific projects.

The following key characteristics of the new SRL scale and assessment are highlighted:

*Unified and universal assessment metrics for project readiness.*
*A set of requirements and conditions ensuring the consistency of assessment results.*
*Consideration of key and additional parameters and criteria for assessing readiness level.*
*Adaptability, reproducibility, comprehensibility, and universality.*

The SRL has been applied in various spheres within the national scientific system:

*A universal and standardized model for assessing the readiness level of scientific projects in the system of R&D grant funding.*
*Qualitative, quantitative, and visual assessment of the readiness level of scientific projects (competitive) for making expert or managerial decisions.*
*The determination of entry points into the grant competition system or the selection of scientific projects.*
*A system for ranking scientific projects based on readiness level parameters and values.*
*Variations and interpretations of the readiness level for specific scientific projects or individual scientific directions.*

Moreover, the readiness level and assessment system will find practical application in scientific policy and science management systems, including the system of expert evaluation and assessment of scientific and research projects.

## 4. Conclusion

The question of the most rational, accurate, and correct assessments and analyses of the quality and effectiveness of a scientist or a scientific organization is crucial for solving the problem of how to build the scientific system, ecosystem, and infrastructure of a state. The role of science in society is not as apparent as, for example, the role of business or agricultural enterprises due to its extraordinary complexity, depth, and, most importantly, the ambiguity of its impact. What useful and specific functions for society and the community do, for example, publication in a scientific journal or the Hirsch index serve?

To answer these and other questions about the importance of science in the life of society, the state, the economy, and the community, a system for evaluating the quality and effectiveness of R&D is necessary. Specifically, it is based on so-called qualimetry and other parallel or sequential theories and practices of quality and effectiveness assessment in general and in science, in particular.

The founders of the term "qualimetry" define the new science as follows: "establishing the principles and regularities of measuring the quality of labor products in general and developing specific methods of such measurement applied to individual types of production" [Asgaldov, 1968, p.34].

By analogy with Asgaldov and other authors, we introduce the concept of "scientific qualimetry" or "epistequalimetry." This can be defined as follows:

*Scientific qualimetry or epistekvalimetry is the science of assessing the quality and effectiveness of R&D.*

Scientific qualimetry or epistequalimetry is a particular case of epistemetriology in the notation of the German-American philosopher Nicholas Rescher [Resher, 2006, p.118].

Accordingly, the new concept of "epistequalimetry" is defined by the following definitions of scientific quality and effectiveness:

*Scientific Quality - a relative measure of compliance with verified and authentic procedures, algorithms, rules, methods, techniques, and ethical and aesthetic norms of scientific processing.*

*Scientific Effectiveness - an absolute measure of the indicators of scientifically informative or material production, interpreted through validation, reliability, reproducibility, and accuracy.*

These definitions do not contradict or correspond to ISO standards and can be formally accepted as universal terms.

Quality and effectiveness are determined based on strict compliance with a certain methodology and technology. In this regard, the level of scientific readiness should determine the quality and effectiveness of scientific work or projects both as a whole and at each level.

Overall, a comprehensive and systematic effort to create a level of scientific readiness and its evaluation system in terms of quality and effectiveness requires solving a number of tasks.

Among these tasks, we can highlight the following main ones:

***Task 1.*** *Complete academic and systematic conceptualization and formalization of the SRL scale for fundamental (FRL), applied (ARL), and innovative projects (IRL).*

The analytical and logical substantiation of the level of scientific readiness, relevance, and limits of applicability to different spheres and branches of scientific directions, including highly integrated, interdisciplinary, and related sciences. This procedure allows the determination of the contours of the universality of the level of scientific readiness regarding the specifics of scientific processing of natural, formal, humanitarian, technical, and other sciences.

***Task 2***. *Identification of Critical Scientific Elements (CSE).*

Critical scientific elements characterize the features, stages, and procedures of scientific processing and scientific work organization in the process of implementing R&D projects. Critical scientific elements allow for a clear gradation and scaling of the stages of scientific processing and, separately, the levels of fundamental (FRL), applied (ARL), and innovative (scientifically intensive) (IRL) projects.

***Task 3.*** *Analytical assessment of each readiness scale in the FRL/ARL/IRL paradigms was performed based on the applicability and adaptation of critical scientific elements (CSE).*

This procedural task will accurately determine the required number of readiness scale levels, their specific content, and semantic character.

***Task 4.*** *Development of a Scientific Maturation Plan (SMP).*

This task involves creating a chain of procedures and stages (master plan) within each scale level to fully comply with a certain level of scientific readiness. That is, each level has sublevels, which determine how much the scientific project has matured for a given level in percentage or relative terms.

***Task 5.*** *Development of a complete protocol and methodology for assessing and rating SRL (scientific readiness levels) based on positional ternary code (integer trits).*

This task will allow for maximum formalization and systematization of the level of scientific readiness of research projects based on visual and numerical representation in the form of a special coding system.

***Task 6.*** *Development of a complete Technical Guide and scientific-technical documentation on the technology, methodology, and processing of the SRL scientific readiness level and ternary code technique.*

This task addresses the issue of content support for the new scale at the expert, professional, training, cognitive, and educational levels.

***Task 7.*** *Creation of a concept and description of the architecture of a software environment for automatic evaluation of the level of readiness of scientific projects based on elements of artificial intelligence (AI).*

Within the framework of this approach, it is necessary to develop a concept and descriptive architecture of a software environment based on a client-server solution for software evaluation of the level of scientific readiness based on elements of artificial intelligence. This software environment will allow documentary information to be obtained through an assessment of the scientific project. The functionality of the software environment can be implemented based on programming algorithms, procedures, and stages of evaluating the level of scientific readiness in terms of quality and effectiveness developed within the framework of this approach.

The solution to these and parallel tasks will yield the following results:

*This paper describes the creation of a universal and standardized methodology for assessing the level of scientific readiness of R&D for practical application in the scientific and grant system of the Republic of Kazakhstan. In particular, management decisions in the national scientific system should be made. New epistemological paradigm - SRL (Scientific Readiness Level)*

*A new system for assessing the level of scientific readiness based on trits (ternary code)*

*This paper provides a full conceptual and epistemological justification of the new methodology for the readiness of the scientific, scientific and technical and innovative projects SRL for R&D (analytical report, scientific and technical information).*

*List, scheme, and scientific-technical documentation of critical scientific elements constituting the level of scientific readiness (analytical report, scientific-technical information)*

*Scientific Maturation Plan (SMP) (analytical report, scientific-technical information)*

*Complete protocol and methodology for assessing and rating SRL (scientific readiness levels) based on positional ternary code (integer trits) (analytical report, scientific-technical information)*

*Concept and descriptive architecture of a software environment (information system) based on a client-server solution for the software evaluation of the level of scientific readiness based on epistemological paradigms and elements of artificial intelligence.*

Within the framework of the level of scientific readiness, the scaling system of scientific readiness can take into account the epistemological coefficient of the significance of a scientific project (SPSC), which evaluates the degree of correlation of the project topic with priority scientific directions and the overall vector of the country's development. This parameter should characterize the level of scientific readiness of the project for perception, synchronicity, and involvement in scientific and technical and socioeconomic development.

The level of scientific readiness should also take into account a new parameter—the Scientific Project Life Cycle (SILC). This parameter characterizes the speed of development of a scientific project. This approach allows the determination of the dynamics of the level of scientific readiness and the scientific prospects of the project in terms of phase transitions and states of rise, maturity, and decline of specific scientific research.

In the context of the general methodology and epistemology, it is also advisable to introduce several types of levels of scientific readiness and their assessment in terms of quality and effectiveness. Namely:

*Individual level of scientific readiness,*
*Project level of scientific readiness*
*National level of scientific readiness*

In this regard, the new system and assessment of the level of scientific readiness reflect a new philosophy and methodology of science in terms of the degree of development and state of scientific fundamental and applied projects.

Overall, the new readiness level scale will allow for an adequate and authentic assessment of the stage of development of a scientific project in purely fundamental and applied projects, as well as the level of commercialization of scientific projects.

In this regard, SRL (scientific readiness levels), when certified and fully standardized, can become a benchmark standard for assessing the level of readiness for R&D. This study is a significant contribution to the development of the philosophy of science, epistemology, scientometrics, science of science, and R&D methodology as a whole.

The level of scientific readiness and the assessment system will find practical application in scientific policy and science management systems (including the system of expertise and assessment of scientific and scientifically intensive projects).

## References

NASA, 2019 - Technology Readiness Level Definitions/nasa.gov. 2019 (дата обращения 11.12.2023)

EU Commission Decision - Technology readiness levels (TRL); Extract from Part 19 - Commission Decision, 2014

V Kazakhstane v khode realizatsii nauchnykh proektov sozdano 140 naukoemkikh proizvodstv, 14.02.2023, https://primeminister.kz/(data obrashcheniya 11.12.2023) (*in Rus*).

Anokhov S.N.. Shkala urovnya tekhnologicheskoi gotovnosti TRL i perspektivy ee modifikatsii. *Business Strategies*. 2022. № 10(11). S. 289-294. (*in Rus*).

Zhumashov, Ye., Azambayev, S., Bisenbaev, A., Yessenzharov, A., & Bulatbayeva, K. Economic efficiency in grant funding evaluations: streamlining knowledge-intensive applications in Kazakhstan//Economic Annals-XXI. 2023. № 201(1-2), pp. 73-83.

Sartori A. V., Gareev A. R., Il'ina N. A., Mantsevich N. M. (2020) Primenenie podkhoda urovnei gotovnosti dlya razlichnykh predmetnykh napravlenii v berezhlivom NIOKR//Ekonomika nauki. 2020. № 6(1–2). S. 118–134 (*in Rus*).

Bisenbaev, A. K. (2023). Optimizatsiya i ratsionalizatsiya podachi zayavok na grantovoe i programmno-tselevoe finansirovanie v Respublike Kazakhstan. *Upravlenie naukoi: teoriya i praktika*. 2023. № 5(3), S. 72-83. (*in Rus*).

Asgal'dov G.G., Glichev A.V.,Krapivenskii Z.N., Kurachenko Yu.P., Panov V.P., Fedorov M.V., Shpektorov D.M. Kvalimetriya – nauka ob izmerenii kachestva produktsii//Standarty i kachestvo, 1968. №1, C. 34-40. (*in Rus*).

Rescher N. Epistemetrics. Cambridge University Press, 2006, p 118